\documentstyle[aps]{revtex}

\begin{document}
\title{Intertwined isospectral potentials in an arbitrary dimension}
\author{\c{S}. Kuru \thanks{%
E.mail:kuru@science.ankara.edu.tr}, A. Te\u{g}men \thanks{%
E.mail:tegmen@science.ankara.edu.tr} and A. Ver\c{c}in \thanks{%
E.mail:vercin@science.ankara.edu.tr}}
\address{Department of Physics \\
Ankara University, Faculty of Sciences,\\
06100, Tando\u gan-Ankara, Turkey}
\maketitle

\begin{abstract}
The method of intertwining with $n$-dimensional ($nD$) linear intertwining
operator ${\cal L}$ is used to construct $nD$ isospectral, stationary
potentials. It has been proven that differential part of ${\cal L}$ is a
series in Euclidean algebra generators. Integrability conditions of the
consistency equations are investigated and the general form of a class of
potentials respecting all these conditions have been specified for each $%
n=2,3,4,5$. The most general forms of $2D$ and $3D$ isospectral potentials
are considered in detail and construction of their hierarchies is exhibited.
The followed approach provides coordinate systems which make it possible to
perform separation of variables and to apply the known methods of
supersymmetric quantum mechanics for $1D$ systems. It has been shown that in
choice of coordinates and ${\cal L}$ there are a number of alternatives
increasing with $n$ that enlarge the set of available potentials. Some
salient features of higher dimensional extension as well as some
applications of the results are presented.
\end{abstract}


\section{Introduction}

The method of intertwining provides a unified approach to constructing
exactly solvable linear and nonlinear problems and their hierarchies in
various fields of physics and mathematics \cite
{Dowker,Anderson2,Matveev,Aratyn,Cannata}. This is closely connected with
the supersymmetric (SUSY) methods such as Darboux's transformation,
Schr\"{o}dinger's factorization, and shape invariant potential concept which
deal with pairs of Hamiltonians having the same energy spectra but different
eigenstates \cite{Junker,Cooper}. In general, the object of the intertwining
is to construct the so called intertwining operator ${\cal L}$ which
performs an intertwining between two given operators of the same type
(differential, integral, matrix, or, operator-valued matrix operator, etc.).
In the context of quantum mechanics ${\cal L}$ is taken to be a linear
differential operator which intertwines two Hamiltonian operators $H_{0}$
and $H_{1}$ such that 
\begin{equation}
{\cal L}H_{0}=H_{1}{\cal L}.
\end{equation}

Two simple and important facts that are at the heart of the usefulness of
this method can be stated as follows; (i) If $\psi ^{0}$ is an eigenfunction
of $H_{0}$ with eigenvalue of $E^{0}$ then $\psi ^{1}={\cal L}\psi ^{0}$ is
an (unnormalized) eigenfunction of $H_{1}$ with the same eigenvalue $E^{0}$.
Hence ${\cal L}$ transforms one solvable problem into another. (ii) When $%
H_{0}$ and $H_{1}$ are Hermitian (on some common function space) ${\cal L}%
^{\dagger }$ intertwines in the other direction $H_{0}{\cal L}^{\dagger }=%
{\cal L}^{\dagger }H_{1}$ and this in turn implies that $[H_{0}, {\cal L}%
^{\dagger }{\cal L}]=0=[{\cal L}{\cal L}^{\dagger },H_{1}]$, where $%
^{\dagger }$ and $[,]$ stand for Hermitian conjugation and commutator.
Therefore , two hidden dynamical symmetry operators of $H_{0}$ and $H_{1}$
are immediately constructed in terms of ${\cal L}$ \cite{Cannata}. These are
dimension and form independent general properties of this method \cite
{Anderson2}. Despite this fact, like the above mentioned SUSY methods,
the intertwining method is mostly studied in the context of one dimensional (%
$1D$) systems where ${\cal L}$ is taken to be first order differential
operator and Hamiltonians are in the standard potential forms. Two
additional properties that arise in that case are that \cite{Anderson3} ;
(i) Every eigenfunction of $H_{0}$ (without regard to boundary conditions or
normalizability) can be used to generate a transformation to a new solvable
problem (see Eq. (20) below). (ii) A direct connection to a SUSY algebra can
be established by constructing a diagonal matrix Hamiltonian $%
H=diag(H_{0},H_{1})$ and two nilpotent supercharges $Q^{+}=(Q^{-})^{\dagger
} $ such that the only nonvanishing element of $Q^{+}$ matrix is $Q_{21}^{+}=%
{\cal L}$. These obey the defining relations of the simplest SUSY algebra 
\begin{eqnarray}
\{Q^{+},Q^{-}\}=H,\quad (Q^{+})^{2}=(Q^{-})^{2}=0,  \nonumber
\end{eqnarray}
which imply $[H, Q^{\pm }]=0$ and emphasize in a compact algebraic form of
the spectral equivalence of two $1D$ systems. In the nomenclature of the
SUSY quantum mechanics ${\cal {L}}$ is known as the supercharge operator and
its zeroth-order (in derivatives) term as the super-potential.

There are important studies in the literature which aim to generalize the
SUSY methods beyond $1D$ problems. These can be classified as (i)
Curved-space approach \cite{Dowker,Anderson2,Anderson1} (for recent studies
see \cite{Veselov}), and (ii) Matrix-Hamiltonian approach \cite
{Cannata,Andrianov1,Andrianov2,Andrianov3}. Both are based on the
intertwining method and they mostly concentrate on extension to two
dimensions.

The application of the first approach to quantum mechanics was motivated by
Ref. \cite{Dowker} which deals with free particle propagation on a Lie group
manifold (see also \cite{Miller}). Later on, this has been advanced to find
the propagator of a free particle moving on an $nD$ sphere \cite{Anderson1}
as well as to solve both ordinary and partial differential equations with
applications to symmetric spaces \cite{Anderson2}. These approaches are
expected to produce solvable $1D$ $n$-particle problems from an $nD$ free
motion via some projection methods like that used in Refs.\cite
{Perelomov,Hoppe}. The second approach, appeared for the first time in Ref. 
\cite{Andrianov1}, performs the extension by preserving the connection with
a SUSY algebra \cite{Andrianov2,Andrianov3}. This inevitably restricts the
consideration to two matrix Hamiltonian such that one of them has
off-diagonal entries. Accordingly, a matrix with elements having higher
order derivative terms participates as the intertwining operator. This
approach establishes the equivalence of two matrix systems but does not
establish spectral equivalence between two scalar Hamiltonians. To improve
it in this regard, an algorithm called the polynomial SUSY in which $%
\{Q^{+},Q^{-}\}$ is a polynomial of the H-matrix was introduced \cite
{Andrianov3}.

The classification given above is by no means exhaustive; for instance one
may find a nice method based on integral intertwining operator in Ref.\cite
{Matveev} (section 2.8) to generate a hierarchy of $2D$ problems. We should
also note that recently the intertwining method has been used for the
non-stationary Schr\"{o}dinger operator\cite{Cannata,Samsonov,Finkel}.

The main purpose of this paper is to extend the intertwining method to an
arbitrary dimension by applying it to a pair of $nD$ systems characterized
by Hamiltonian operators of potential form 
\begin{equation}
H_{i}=-\nabla ^{2}+V_{i},\quad i=0,1,
\end{equation}
where the potentials $V_{i}$ and eigenvalues of $H_{i}$ are expressed in
terms of $2m/\hbar ^{2}$ and $\nabla ^{2}=\sum_{j=1}^{n}\partial _{j}^{2}$
is the Laplace's operator of {\bf $R^{n}$}. We shall use the Cartesian
coordinates $\{x_{k};k=1,...,n\}$, the convention $\partial _{k}\equiv
\partial /\partial x_{k} $ and the abbreviation $V_{i}\equiv
V_{i}(x_{1},...,x_{n})$ throughout the paper. We purpose the ansatz that $%
{\cal L}$ is the most general first-order linear operator 
\begin{equation}
{\cal L}=L_{0}+L_{d}=L_{0}+\sum_{k=1}^{n}L_{k}\partial _{k}
\end{equation}
where $L_{0},L_{k}$ are some functions of $\{x_{k};k=1,...,n\}$ which
together with $V_{i}$ are to be determined from consistency equations of Eq.
(1). In terms of the vector field $\vec{L}=(L_{1},...,L_{n})$ and $nD$
gradient operator $\vec{\nabla}$ the operator $L_{d}=\sum_{k=1}^{n}L_{k}%
\partial _{k}$ will be usually written as $L_{d}=\vec{L}\cdot \vec{\nabla}$,
where ``$\cdot$'' denotes the usual Euclidean inner product.

In the next section by solving the first $n(n+1)/2$ consistency equations we
will show that the operator $L_{d}$ is a series in generators of the
Euclidean algebra in $n$-dimension. There remain $n+1$ consistency equations
which consist of $n$ linear and $1$ non-linear partial differential
equations. Some particular solutions of these equations for an arbitrary $n$
are presented in section III. In section IV we take up the integrability
conditions of the remaining $n$ linear equations in the context of the
Frobenius integrability theory \cite{Flanders}. General forms of the
potentials respecting all integrability conditions for $n=2,3,4,5$ are
obtained in section V. A detailed investigation of $2D$ and $3D$ isospectral
potentials are given in sections VI, VII VII where we also exhibit how to
generate hierarchies of potentials.

\section{Intertwining in {\sl n} Dimension: Euclidean Algebra}

In view of (2) and (3) the intertwining relation (1) can be written as 
\begin{equation}
\lbrack \nabla ^{2},L_{d}]=-[\nabla ^{2},L_{0}]+[V_{0},L_{d}]+P{\cal L},
\end{equation}
where $P=V_{1}-V_{0}$. At a glimpse of the right hand side of Eq. (4) and 
\begin{eqnarray}
\lbrack \nabla ^{2},L_{d}]&=&\sum_{j,k}(\partial _{j}^{2}L_{k})\partial
_{k}+2\sum_{j}(\partial _{j}L_{j})\partial _{j}^{2}+2\sum_{j<k}(\partial
_{j}L_{k}+\partial _{k}L_{j})\partial _{j}\partial _{k}, \\
\lbrack \nabla ^{2},L_{0}]&=&(\nabla ^{2}L_{0})+2\sum_{j}(\partial
_{j}L_{0})\partial _{j}, \\
\lbrack V_{0},L_{d}]&=&-(L_{d}V_{0})=-\sum_{j} L_{j}(\partial _{j}V_{0}),
\end{eqnarray}
we see that the second order derivatives in Eq. (4) come, together with some
first order derivatives, only from $[\nabla ^{2},L_{d}]$. Therefore by
setting their coefficients to zero we obtain two sets of consistency
equations: 
\begin{equation}
\partial _{j}L_{j}=0,\quad j=1,...,n;\qquad\partial _{j}L_{k}+\partial
_{k}L_{j}=0,\quad j<k=2,...,n.
\end{equation}
The first set gives $L_{j}=a_{j}+f_{j}(x)$, where $a_{j}$'s are constants
and $f_{j}(x)$ depends on all of $x_{k}$'s except $x_{j}$. The second set
determines $f_{j}$ as $f_{j}=\sum_{k}c_{jk}x_{k}$ where $c_{jk}$'s are all
constants and antisymmetric in $j$ and $k:c_{jk}+c_{kj}=0$. Hence 
\begin{equation}
L_{j}=a_{j}+\sum_{k}c_{jk}x_{k}.
\end{equation}

These solutions make the first order derivative terms at the right hand side
of (5) vanish so that $[\nabla ^{2},L_{d}]=0$. As a result of this the
intertwining relation (4) simplifies to 
\begin{equation}
\lbrack \nabla ^{2},L_{0}]=[V_{0},L_{d}]+P(L_{0}+L_{d}).
\end{equation}
From (6), (7) and (10) we get, by equating the coefficients of the first and
zeroth powers of derivatives 
\begin{eqnarray}
2\partial _{j}L_{0} &=&PL_{j};\quad j=1, 2,..., n, \\
(-\nabla ^{2}+P)L_{0} &=&(L_{d}V_{0}).
\end{eqnarray}
These $n+1$ equations constitute a reduced form of the consistency
conditions for three unknown functions $L_{0}, V_{0}$ and $V_{1}$. While Eq.
(12) is non-linear, Eqs. (11) are linear since all components of $L_{d}$
have been found.

Eq. (12) can be considered in the following way. By virtue of 
\begin{equation}
\partial_{j}L_{k}=c_{kj},
\end{equation}
Eqs. (11) imply that 
\begin{equation}
\nabla^{2}L_{0}=\frac{1}{2}(L_{d}P).
\end{equation}
Combining this with (12) we arrive at 
\begin{equation}
L_{0}P=\frac{1}{2}L_{d}(V_{1}+V_{0}).
\end{equation}
which can be used instead of Eq. (12).

By defining 
\begin{equation}
T_{j}=\partial _{j},\qquad L_{jk}=x_{k}\partial _{j}-x_{j}\partial _{k},
\end{equation}
and using (9) $L_{d}$ can be written as 
\begin{equation}
L_{d}=\sum_{j}a_{j}T_{j}+\sum_{j<k}c_{jk}L_{jk}.
\end{equation}
The generators $T_{j}$'s and $L_{jk}$'s obey the following commutation
relations 
\begin{eqnarray}
\lbrack T_{j},T_{k}] &=&0,  \nonumber \\
\lbrack T_{j},L_{km}] &=& \delta _{jm}T_{k}-\delta_{jk}T_{m}, \\
\lbrack L_{jk},L_{\ell m}] &=& \delta _{jm}L_{\ell k} -\delta
_{j\ell}L_{mk}+\delta_{k\ell}L_{mj}-\delta _{km}L_{\ell j}.  \nonumber
\end{eqnarray}
These are the defining relations of $n(n+1)/2$ dimensional Euclidean algebra 
$e(n)$, also known as the algebra of rigid motion denoted by $iso(n)$ \cite
{Miller,Gilmore}. $n$ translational generators $T_{j}$'s form the invariant
abelian subalgebra $t(n)$ and $n(n-1)/2$ rotational generators $L_{jk}$'s
form the semisimple subalgebra $so(n)$. As is well known $e(n)$ is
semi-direct sum of $t(n)$ and $so(n)$ and $\sum T_{j}^{2}=\nabla ^{2}$ is a
Casimir operator of $e(n)$.

Now, we shall show that the above analysis includes and naturally
generalizes the well known $1D$ case. It is evident that for $n=1$ we have $%
{\cal L}=L_{0}+\partial _{x}$ and $P=2L_{0}^{\prime }(x)$, where we take $%
x\equiv x_{1},a_{1}=1$ and we use the prime(s) to denote differentiation(s)
with respect to the argument (when there is no risk of confusion the
argument will be suppressed). In that case Eq. (15) yields $\partial
_{x}(V_{0}+L_{0}^{\prime }-L_{0}^{2})=0$ from which we recover the
well-known forms of the $1D$ partner potentials: 
\begin{equation}
V_{0}=L_{0}^{2}-L_{0}^{\prime }+b,\qquad V_{1}=L_{0}^{2}+L_{0}^{\prime }+b.
\end{equation}
It is a standard procedure of $1D$ SUSY quantum mechanics to take the
constant $b$ and $L_{0}$ as $b=\lambda _{1}$ and $L_{0}(x)=-\partial
_{x}[\ln \phi _{1}(x)]$. When these are substituted into the first equation
of (19) we obtain : $-\phi _{1}^{\prime \prime }(x)+V_{0}\phi
_{1}(x)=\lambda _{1}\phi _{1}(x)$, that is, $\phi _{1}(x)$ is the
eigenfunction of the Schr\"{o}dinger's equation $-\phi ^{\prime \prime
}(x)+V_{0}\phi (x)=\lambda \phi (x)$ corresponding to the eigenvalue $%
\lambda =\lambda _{1}$. Therefore the Schr\"{o}dinger's equation remains
covariant under the Darboux's transformations 
\begin{equation}
(\phi ,V_{0})\rightarrow (L\phi =\phi ^{\prime }-[\ln \phi _{1}]^{\prime
}\phi ,\quad V_{1}=V_{0}-2[\ln \phi _{1}]^{\prime \prime }).
\end{equation}
Obviously, instead of $\phi _{1}$, any other fixed eigenfunction can be used
to generate a transformation to another new potential $V_{1}$. It is this
fact which allows us to apply the Darboux's transformations successively and
to construct a hierarchy of potentials for a given $V_{0}$.

We conclude this section by saying that for $n\geq 1$ the differential part
of the intertwining operator is a series in generators of $e(n)$. In saying
that we have identified the algebra generated by $\partial _{x}$ with $e(1)$%
. A related result is that intertwined potentials have symmetry generators
differential part of which are quadratic in the generator of $e(n)$, that
is, they belong to universal enveloping algebra of $e(n)$. From now on we
assume that $a_{j}$'s and $c_{jk}$'s are real constants.

\section{Applications}

Before proceeding further we consider some particular cases of Eqs. (11) and
(12).

When $P=0$ Eqs. (11) and (15) give $L_{0}=$constant and $(L_{d}V_{0})=0$. In
view of Eq. (1) these imply, as an expected result, that ${\cal L}$ is a
symmetry generator of $H_{0}=H_{1}:[H_{0}, {\cal L}]=0$.

Next we take $P$ to be a constant such that $P=p_{0}\neq 0$. In that case
the integrability conditions $\partial _{j}\partial _{k}L_{0}=\partial
_{k}\partial _{j}L_{0}$ of Eqs. (11) require that $c_{jk}=0$ for all $j,k$
which lead to $2L_{0}=p_{0}\vec{a}\cdot \vec{r}+2b$, where $\vec{r}%
=(x_{1},...,x_{n})$ is the position vector and $\vec{a}$ represents the
constant vector $\vec{a}=(a_{1},...,a_{n})$. Taking the constant $b$ as $b=%
\vec{a}\cdot \vec{b}$ we get from (15) 
\[
\vec{a}\cdot (p_{0}^{2}\vec{r}+2p_{0}\vec{b}-2\vec{\nabla}V_{0})=0, 
\]
which is solved by 
\begin{equation}
V_{0}=\frac{1}{4}p_{0}^{2}r^{2}+p_{0}\vec{b}\cdot \vec{r}+g(x),
\end{equation}
where $\vec{b}$ is a constant vector, $r^{2}=\sum_{j}x_{j}^{2}$ and $%
g(x)\equiv g(x_{1},...,x_{n})$ is any differentiable function subjected to
the constraint $\vec{a}\cdot \vec{\nabla}g(x)=0$. One may take 
\begin{equation}
g(x)=g(\vec{b}_{(1)}\cdot \vec{r},\dots ,\vec{b}_{(n-1)}\cdot \vec{r}),
\end{equation}
such that $\vec{b}_{(j)}$'s are linearly independent vectors perpendicular
to $\vec{a}$. Different choices of $g$ define different systems which accept 
\begin{equation}
{\cal L}^{\dagger }{\cal L}=-(\vec{a}\cdot \vec{\nabla})^{2}+[\vec{a}\cdot (%
\frac{1}{2}p_{0}\vec{r}+\vec{b})]^{2}-\frac{1}{2}p_{0}a^{2},
\end{equation}
as a common symmetry generator. Accordingly 
\begin{equation}
{\cal L}{\cal L}^{\dagger }=-(\vec{a}\cdot \vec{\nabla})^{2}+[\vec{a}\cdot (%
\frac{1}{2}p_{0}\vec{r}+\vec{b})]^{2}+\frac{1}{2}p_{0}a^{2},
\end{equation}
is a common symmetry generator for $V_{1}=V_{0}+p_{0}$. These also imply
that ${\cal L}/a$ and ${\cal L}^{\dagger }/a$ are a pair of ladder operators
for $H_{0}$: 
\begin{eqnarray}
[H_{0}, {\cal L}]=-p_{0}{\cal L},\quad [H_{0}, {\cal L}^{\dagger}]=p_{0}%
{\cal L}^{\dagger}, \quad [{\cal L}, {\cal L}^{\dagger }]=p_{0}a^{2}. 
\nonumber
\end{eqnarray}
As a result of these we recover the existence of harmonic oscillator like
spectrum in the spectrum of a class of $nD$ systems described by $H_{0}$
which contains many parameters and an arbitrary function.

Now we set all of $c_{jk}$'s to zero. From (11) and (15) we get $%
L_{0}=f(\zeta)$ and $P=f^{\prime}(\zeta )$ where $f$ is an arbitrary
differentiable function of $\zeta =\vec{a}\cdot \vec{r}/2$. Defining 
\begin{equation}
V_{\pm}=\frac{1}{a^{2}}f^{2}(\zeta)\pm \frac{1}{2}f^{\prime}(\zeta)
\end{equation}
we obtain, by virtue of (11) and (15) 
\begin{equation}
V_{0}=\frac{1}{2}g(x)+V_{-}; \qquad V_{1}=\frac{1}{2}g(x)+V_{+},
\end{equation}
where $g(x)$ may be taken as in (22). Observing that $V_{\pm}$ are form
equivalent to (19) we can say that all the known techniques of $1D$ SUSY
quantum mechanics can equally well be used in this case. For this
application the intertwining operator is ${\cal L}=f(\zeta)+\vec{a}\cdot\vec{%
\nabla}$ and the symmetry generators are 
\begin{equation}
{\cal L}{\cal L}^{\dagger}=a^{2}V_{+}-(\vec{a}\cdot \vec{\nabla})^{2},\qquad 
{\cal L}^{\dagger}{\cal L}=a^{2}V_{-}-(\vec{a}\cdot \vec{\nabla})^{2}.
\end{equation}

\section{Integrability Conditions}

In this section we concentrate on the integrability conditions of $n$ linear
equations given by (11). It turns out that once these conditions are well
understood all the consistency equations can be tackled more easily.

By considering $L_{0}$ as the $(n+1)$-th coordinate $x_{n+1}\equiv L_{0}$ of 
{\bf $R^{n+1}$} and $P$ as a function defined on it we introduce the 1-form 
\begin{equation}
\Omega =dL_{0}-\frac{1}{2}P\Gamma ,
\end{equation}
on {\bf $R^{n+1}$}. Here $d$ stands for the exterior derivative and $\Gamma $
denotes the 1-form 
\begin{equation}
\Gamma =\sum_{j=1}^{n}L_{j}dx_{j},
\end{equation}
on {\bf $R^{n}$}. Now $n$ linear equations given by (11) can be expressed as
a single Pfaffian equation $\Omega =0$. In the Frobenius theory,
integrability of this Pfaffian equation amounts to being able to find a
positive valued integrating factor $f$ and a function $g$ such that $\Omega
=fdg$ \cite{Flanders}. If this is possible then $\Omega =0$ and $dg=0$ are
equivalent Pfaffian equations and the solution (integral surface) of $\Omega
=0$ is the hypersurface $g=$constant. According to the Frobenius theorem a
necessary and sufficient condition for the existence of functions $g$ and $f$
is the fulfillment of the so called Frobenius condition: 
\begin{equation}
\Omega \wedge d\Omega =0,
\end{equation}
where $\wedge $ denotes the usual exterior product.

From (28) and (29) we have 
\begin{eqnarray}
d\Omega=-\frac{1}{2}[(\partial_{n+1}P)dL_{0}\wedge \Gamma
+\sum_{j=1}^{n} (\partial_{j}P)dx_{j}\wedge \Gamma +Pd\Gamma],  \nonumber
\end{eqnarray}
and therefore 
\begin{equation}
\Omega \wedge d\Omega =-\frac{1}{2}[dL_{0}\wedge d(P\Gamma )-\frac{1}{2}
P^{2}\Gamma \wedge d\Gamma ],
\end{equation}
where $d$ in $d(P\Gamma)$ and $d\Gamma$ stands for the exterior derivative
of $R^{n}$. The Frobenius conditions (30) is therefore equivalent to the
following two conditions 
\begin{eqnarray}
d(P\Gamma )&=&0, \\
\Gamma\wedge d\Gamma &=&0,
\end{eqnarray}
provided that $P\neq 0$. Since both of these conditions are valid in $R^{n}$%
, $P$ is defined on $R^{n}$.

The condition (32) gives $n(n-1)/2$ equations 
\begin{equation}
K_{jk}P=-2c_{jk}P,
\end{equation}
where 
\begin{equation}
K_{jk}=L_{j}\partial _{k}-L_{k}\partial _{j}.
\end{equation}
Observe that Eq. (34) can also be obtained from $\partial
_{j}\partial_{k}L_{0}=\partial _{k}\partial _{j}L_{0}$ and in deriving it we
have used Eq. (13). The condition (33) could also be inferred from Eq. (32)
upon exterior multiplication of $dP\wedge \Gamma+Pd\Gamma=0 $ by $\Gamma$.
It leads to $n(n-1)(n-2)/6$ equations: 
\begin{equation}
L_{[j}c_{k\ell ]}=0,
\end{equation}
where $j<k<\ell \leq n$ and the square bracket $[\quad ]$ enclosing the
subindexes means anti-symmetrization. Eq. (36) shows that any three of $%
L_{j} $'s are linearly dependent, that is 
\begin{equation}
L_{j}c_{k\ell }+L_{k}c_{\ell j}+L_{\ell }c_{jk}=0.
\end{equation}
Making use of (9) this can be written as 
\begin{equation}
a_{[j}c_{k\ell ]}=\sum_{m}x_{m}c_{m[j}c_{k\ell ]}.
\end{equation}
This gives nothing in the case of $n=2$ because $\Gamma \wedge d\Gamma $ is
a 3-form and therefore identically vanishes on {\bf $R^{2}$}.

By a simple reasoning making use of the anti-symmetry of $c_{jk}$'s we see
that for $n=3$ the right hand side of Eq. (38) vanishes identically and a
single condition 
\begin{equation}
\vec{L}\cdot \vec{c}=\vec{a}\cdot \vec{c}=0
\end{equation}
results. Here we have made use of the fact that in the case of $n=3$ we have 
\begin{equation}
\vec{L}=\vec{a}+\vec{r}{\bf \times}\vec{c}
\end{equation}
where $\vec{c}=(c_{1},c_{2},c_{3})=(c_{23},c_{31},c_{12})$ and ``${\bf \times%
}$'' stands for the usual cross product of $R^{3}$. For $n\geq 4$ more care
is needed. It is not hard to check that $c_{m[m}c_{k\ell ]}=0$ for any $n$
and hence for $n=4$ the terms $c_{1[j}c_{k\ell ]},...,c_{4[j}c_{k\ell ]}$
are equal to each other up to a sign $``-"$. These imply that in the case of 
$n=4$, Eqs. (38) restrict all the coordinates to some constant values. But,
as is evident from Eqs. (38), at the expense of constraining the form of $L$
we can get rid of all these coordinate restrictions by imposing the
conditions 
\begin{eqnarray}
a_{[j}c_{k\ell ]} &=&0\quad ;\qquad j<k<\ell , \\
c_{m[j}c_{k\ell ]} &=&0\quad ;\quad m=1,...,n.
\end{eqnarray}

As is mentioned above in the case of $n=4$ Eqs. (42) give only one condition 
\[
c_{12}c_{34}+c_{13}c_{42}+c_{14}c_{23}=0, 
\]
and Eqs. (41) give conditions which reduce the total number of parameters.
To see this more concretely we define the following four vectors 
\begin{eqnarray}
\vec{c}_{(1)} &=&(0,c_{34},-c_{24},c_{23}),\quad \vec{c}%
_{(2)}=(c_{34},0,c_{41},-c_{31}),  \nonumber \\
\vec{c}_{(3)} &=&(c_{24},c_{41},0,c_{12}),\quad \vec{c}%
_{(4)}=(c_{23},c_{31},c_{12},0).  \nonumber
\end{eqnarray}
Now Eqs. (41) can be rewritten as $\vec{a}\cdot \vec{c}_{(j)}=0,j=1,2,3,4$.
It is easy to check that, in view of Eq. (42), the determinant of the matrix
formed by the components of the vectors $\vec{c}_{(j)}$'s has rank two.
Therefore Eqs. (41) provide two of $a_{j}$'s as free parameters, or, for a
given $\vec{a}$ two constraints for $c_{jk}$'s. By taking into account also
(42) we get seven free parameters: five $c_{jk}$'s and two $a_{j}$'s, or,
three $c_{jk}$'s and four $a_{j}$'s. These can be chosen in many different
ways. Moreover, one can also chose a lesser number of parameters without
destroying the integrability conditions.

For the number of conditions implied by (41-42) exceeds the number of
parameters the investigation is getting harder and harder for $n\geq 5$.
But, in the case of $n=5$ one can keep again $5$ of $c_{jk}$'s as free
parameters by setting all $a_{j}$'s to zero. In that case Eqs. (41)
disappear and Eqs. (42) give $5$ constraints which reduce the number of $%
c_{jk}$'s from $10$ to $5$. Note also that, as has been done in section III,
for $n\geq 2$ one can always set all $c_{jk}$'s to zero and keep $n$ $a_{j}$%
's as parameters. In such a case the condition (36) completely disappears.

These remarks imply an important property of the intertwining method in
higher dimensions; due to integrability conditions there are a number of
choices in specifying ${\cal L}$. Evidently this fact enriches the set of
intertwined potentials (see the Table I in the case of $n=3$). In the next
section by taking $c_{jk}\neq 0$ for at least a pair of $j, k$, we carry out
an investigation which will enable us to find out the general forms of a
class of potentials for $n=2,3,4,5$ endowed with mentioned richness for $%
n\geq 3$.

\section{General Form of Potentials}

By making use of Eqs. (13), (35) and (37) one can easily verify the
following relations 
\begin{eqnarray}
\partial_{m}(\frac{L_{i}}{L_{j}})&=&\frac{c_{ij}}{L_{j}^{2}}L_{m} , \\
K_{mn}(\frac{L_{i}}{L_{j}})&=&0 , \\
K_{mn}(\frac{1}{L_{j}^{k}})&=&-k\frac{c_{mn}}{L_{j}^{k}}, \\
\vec{L}\cdot \vec{\nabla}(\frac{L_{i}}{L_{j}})&=& c_{ij}\frac{L^{2}}{%
L_{j}^{2}}, \\
\vec{L}\cdot \vec{\nabla}g(L^{2})&=&
(2\sum_{ij}L_{i}L_{j}c_{ij})g^{\prime}(L^{2})=0.
\end{eqnarray}
In Eq. (47) $g$ is an arbitrary function of $L^{2}=\sum_{j}L_{j}^{2}$.
Comparing Eqs. (43) and (11) we see that the general form of $L_{0}$ is 
\begin{equation}
L_{0}=f(\frac{L_{i}}{L_{j}}),
\end{equation}
provided that $c_{ij}\neq 0$. Then from any of Eqs. (11) $P$ is found to be 
\begin{equation}
P=\frac{2c_{ij}}{L_{j}^{2}}f^{\prime}(\eta ),
\end{equation}
where $\eta=L_{i}/L_{j}$. Fortunately, Eqs. (44) and (45) imply that the
solution (49) respects all the integrability conditions given by (34).

The only equation that remained unsolved is Eq. (15) which is now as follows 
\begin{equation}
\vec{L}\cdot \vec{\nabla}(V_{1}+V_{0})= \frac{2c_{ij}}{L_{j}^{2}}%
\partial_{\eta}[f^{2}(\eta )].
\end{equation}
From (46) and (47) it is evident that the general solution of this equation
is of the form 
\begin{equation}
V_{1}+V_{0}= h+ 2\frac{f^{2}(\eta)}{L^{2}},
\end{equation}
where $2f^{2}(\eta)/L^{2}$ accounts for the right hand side of (50) and $h$
is the general solution of the homogeneous equation 
\begin{equation}
\vec{L}\cdot \vec{\nabla}h=0.
\end{equation}
Hence, the general forms of $V_{0}$ and $V_{1}$ are, by combining (49) and
(51) 
\begin{eqnarray}
V_{0}&=&\frac{1}{2}h+\frac{V_{-}}{L^{2}}, \\
V_{1}&=&\frac{1}{2}h+\frac{V_{+}}{L^{2}},
\end{eqnarray}
where 
\begin{equation}
V_{\pm}= f^{2}(\eta)\pm c_{ij}\frac{L^{2}}{L_{j}^{2}}f^{\prime}(\eta).
\end{equation}

As a result, the number of consistency equations has been reduced from $%
(n+1)(n+2)/2$ (the sum of the number of Eqs. (8),(11) and (12)) to $1$,
i.e., to Eq. (52). Geometrically, Eq. (52) means that at each point of the
surface $h=$constant, $\vec{L}$ always lies on the local tangent space.
Equivalently, $\vec{L}$ is always perpendicular to the (classical) force
field determined by $\vec{\nabla}h$. On the other hand, from group
theoretical point of view Eq. (52) means that the common part of the
intertwined potentials is invariant under the action of the Euclidean group $%
E(n)$, i.e., $e^{L_{d}}h=h$. For all these statements and the integrability
conditions are dimension-dependent $h$ must be determined in each case
separately. The rest of the paper is devoted to a detailed investigation of $%
n=2 $ and $n=3$ cases.

As our investigation for an arbitrary dimension is completed two remarks are
in order. (i) The above analysis enables us to write down a class of $nD$
isospectral potentials provided that at least one of $c_{jk}$'s is different
from zero. For instance, if only $c_{jk}\neq 0$ then Eqs. (37) imply that $%
L_{m}=0$ for $m\neq j,k$ and Eqs. (11) require $L_{0}$ to depend only on $%
x_{j}$ and $x_{k}$. In such a case, after defining $\eta =L_{j}/L_{k}$ it
remains to solve Eq. (52) to find suitable $n-1$ coordinate functions. (ii)
When the number of non-zero $c_{jk}$'s is greater than one there are a
number of choices (at most $n(n-1)/2$) for $\eta $. But, from Eq. (37) we
see that these are all functionally dependent to each other. For example, in
the case of $n=3$ we have three choices $\eta =L_{1}/L_{2},\eta
_{2}=L_{1}/L_{3},\eta _{3}=L_{2}/L_{3}$ which obey the following relations 
\[
\eta _{3}=\eta _{2}/\eta ,\qquad \eta _{2}c_{23}+\eta _{3}c_{31}=-c_{12}. 
\]
Instead of $\eta _{i}$ one may choose one of the variables $\alpha
_{i}=L_{i}/\vec{r}\cdot \vec{L}=L_{i}/\vec{r}\cdot \vec{a}$, or for $%
n=3,\sigma _{i}=L_{i}/(\vec{c}{\bf \times }\vec{L})_{i}$. It is easy to
verify that each of these satisfies relations similar to Eqs. (43-44) and
(46) and enables us to express $L_{0},P,V_{\pm }$ in terms of them. This
freedom in the choice of coordinates once again manifests the largeness of
the set of intertwined potentials. But, we should emphasize that these are
all functionally dependent since the differential of any variable obeying
(43) is proportional to $\Gamma =\vec{L}\cdot d\vec{r}$ and therefore $d\eta
_{i}\wedge d\alpha _{i}=0$, etc. This also proves that as long as first
order intertwining is concerned $V_{\pm }$ depend only on one variable.

\section{2D Isospectral Potentials}

In two dimension we have $L_{1}=(a_{1}+cy)$ and $L_{2}=(a_{2}-cx)$, where $%
c=c_{12}, x=x_{1}, y=x_{2}$. From Eq. (47) we see that, in terms of 
\begin{equation}
\kappa=[L_{1}^{2}+L_{2}^{2}]^{1/2}=[(a_{1}+cy)^{2}+(a_{2}-cx)^{2}]^{1/2}
\end{equation}
the general solution of Eq. (52) is $h=h(\kappa)$, where $h$ is an arbitrary
differentiable function. Taking $\eta=L_{1}/L_{2}$ and noting that $%
L^{2}/L_{2}^{2}=1+\eta^{2}$, by Eqs. (53-55) the general forms of the $2D$
isospectral potentials are found to be 
\begin{equation}
V_{0}=\frac{1}{2}h(\kappa)+\frac{V_{-}}{\kappa^{2}},\quad V_{1}=\frac{1}{2}%
h(\kappa)+\frac{V_{+}}{\kappa^{2}},
\end{equation}
where 
\begin{equation}
V_{\pm}= f^{2}(\eta)\pm c(1+\eta^{2})f^{\prime}(\eta).
\end{equation}
In that case the intertwining operator is 
\begin{eqnarray}
{\cal L}=f(\eta)+ (a_{1}+cy)\partial_{x}+(a_{2}-cx)\partial_{y}
=f(\eta)+c(1+\eta^{2})\partial_{\eta}.
\end{eqnarray}

As is well known, for a $2D$ stationary system the existence of a symmetry
generator means that the system is completely integrable in the Liouville
sense. Recalling that ${\cal L}^{\dagger }{\cal L}$ and ${\cal L}{\cal L}%
^{\dagger }$ are symmetry generators of $H_{0}$ and $H_{1}$, the potentials
given by (57) are the most general forms of $2D$ integrable potentials which
can be intertwined by a first order operator.

We shall now present some examples in which for some simple forms of $V_{-}$
we consider the Riccati's equation (58) for dependent variable $f$ and by
solving it we construct the corresponding potentials. As the simplest case
we take $V_{0}=0$. This may happen in two different cases; (i) $h=0, V_{-}=0$%
, and (ii) $h=-2b/\kappa^{2}, V_{-}=b$, where $b$ is a constant. In these
cases (58) is a separable equation of the form 
\begin{equation}
f^{2}-c(1+\eta^{2})f^{\prime}=b,
\end{equation}
which has the general solution 
\begin{equation}
f=(-b)^{1/2}\tan[\frac{(-b)^{1/2}}{c}(\tan^{-1}\eta-b_{1})]
\end{equation}
for $b<0$. This should be read as $f=b^{1/2}\tanh[(b^{1/2}/c)(b_{1}-%
\tan^{-1}\eta)]$ for $b>0$ and as $f=c(b_{1}-\tan^{-1}\eta)^{-1}$ for $b=0$,
where $b_{1}$ is an integration constant. From (58) we have 
\begin{equation}
V_{1}=2c^{2}[\kappa (b_{1}-\tan^{-1}\eta)]^{-2},
\end{equation}
for the case (i) and 
\begin{eqnarray}
V_{1}=-2b\{ \kappa \cos[\frac{(-b)^{1/2}}{c}(\tan^{-1}\eta-b_{1})]\}^{-2};
\quad V_{1}=-2b\{ \kappa \cosh[\frac{b^{1/2}}{c}(\tan^{-1}\eta-b_{1})]
\}^{-2},
\end{eqnarray}
for the case (ii) corresponding to $b<0$ and $b>0$ respectively. As a result
we have found a two parameter family of $2D$ potentials that are intertwined
to $2D$ free motion. Note that for $b=-c^{2}, b_{1}=0$ we have $f=c\eta $
and 
\begin{equation}
V_{1}=2c^{2}\frac{\eta^{2}+1}{\kappa^{2}}= \frac{2c^{2}}{(a_{2}-cx)^{2}}.
\end{equation}

As another example, taking $V_{-}=b=-c^{2}$ and $h=(2c^{2}/\kappa
^{2})+2g(\kappa )$ in (57) leads us to the partner potentials 
\begin{equation}
V_{0}=g(\kappa ),\quad V_{1}=g(\kappa )+2c^{2}\frac{1+\eta ^{2}}{\kappa ^{2}}
\end{equation}
for $f=c\eta $. In particular, for $g(\kappa )=\kappa ^{2}$, $H_{0}$
represents a $2D$ isotropic displaced harmonic oscillator and $H_{1}$ a $2D$
Calogero's type system for which 
\begin{equation}
V_{1}=\frac{2c^{2}}{(a_{2}-cx)^{2}}+(a_{2}-cx)^{2}+(a_{1}+cy)^{2}.
\end{equation}
In that case for any choice of $g(\kappa )$ we have ${\cal L}=c[\eta +(\eta
^{2}+1)\partial _{\eta }]$. This explicitly shows that two different
families of potentials, such as that given by (65) can be intertwined by the
same ${\cal L}$. This is an important property that we do not have in one
dimension. It is evident that this arises from the separability of the
problem that we shall analyze in the next section. It is also worth
mentioning that after a simple affine transformation of the coordinates and
a restriction on $c^{2}$ one can easily recognize (66) as one of the four
superintegrable the Smorodinsky-Winternitz $2D$ potentials \cite{Evans}. The
above particular example shows that this potential is intertwined to the
harmonic oscillator and one of its symmetry generators is immediately
obtained as ${\cal L}{\cal L}^{\dagger }$.

\section{Separation of Variables and Hierarchy of $2D$ Potentials}

The above analysis suggests the variables ($\kappa, \eta$) as a new
coordinate system. This is a kind of the orthogonal polar coordinate system
with displaced center in which we have 
\begin{equation}
\nabla ^{2} = \frac{c^{2}}{\kappa^{2}}\{ \kappa \partial _{\kappa }(\kappa
\partial _{\kappa }) +(1+\eta^{2})\partial
_{\eta}[(1+\eta^{2})\partial_{\eta}]\}.
\end{equation}
This implies that the eigenvalue equations of $H_{i}$ accept the separation
of variables in terms of ($\kappa, \eta $). In fact, this can be carried out
in an easier way by introducing the coordinates 
\begin{equation}
\rho=\frac{1}{c}\ln\kappa,\qquad \xi=\frac{1}{c}\tan^{-1}\eta.
\end{equation}
From (59) and (67) we get 
\begin{equation}
{\cal L}=f(\xi )+\partial _{\xi },
\end{equation}
and $\nabla ^{2}=e^{-2c\rho } (\partial_{\rho }^{2}+\partial_{\xi }^{2})$.
By defining 
\begin{eqnarray}
H_{\rho }=-\partial _{\rho }^{2}+\frac{1}{2}e^{2c\rho }h(\rho ),\quad H_{\pm
}=-\partial _{\xi }^{2}+V_{\pm }(\xi ).  \nonumber
\end{eqnarray}
and 
\begin{equation}
V_{\pm }=f^{2}(\xi )\pm f^{\prime }(\xi ).
\end{equation}
the Hamiltonians can be written as 
\begin{eqnarray}
H_{0}=e^{-2c\rho }(H_{\rho }+H_{-}),\quad H_{1}=e^{-2c\rho }(H_{\rho
}+H_{+}).
\end{eqnarray}

If we take $\psi ^{0}(\rho, \xi)=R(\rho )U^{0}(\xi )$ the eigenvalue
equation $H_{0}\psi ^{0}(\rho, \xi )=E^{0}\psi ^{0}(\rho, \xi)$ separates as
follow 
\begin{eqnarray}
H_{-}U^{0}(\xi ) &=&MU^{0}(\xi ), \\
(H_{\rho }-E^{0}e^{2c\rho })R(\rho ) &=&-MR(\rho ),
\end{eqnarray}
where $M$ is the separation constant. For given $E^{0}$ $\rho $-equation for 
$H_{1}$ is the same as Eq. (73), but $\xi $-equation is $H_{-}U^{1}(\xi
)=MU^{1}(\xi )$. ${\cal L}$ given by (69) intertwines only solutions of $%
H_{-}$ to that of $H_{+}$ by $U^{1}(\xi )={\cal L}U^{0}(\xi )$.

We shall now briefly describe how to generate a hierarchy of $2D$
isospectral potentials.

Taking $f(\xi) =-\phi^{\prime}(\xi)/\phi (\xi)$ in Eq. (70) yields $%
V_{-}(\xi)=\phi^{\prime \prime}(\xi)/\phi(\xi)$. This is the same as Eq.
(72) for $M=0$. Therefore, each solution of (72) with $M=0$ can be used to
generate a transformation to a new problem with potential $V_{1}$. In fact,
by keeping analogy with $1D$ SUSY methods we can do more than that. For this
purpose let us take 
\begin{equation}
V_{-}(\xi )={\cal V}(\xi )-{\cal E}_{n}, \qquad f(\xi)=-\frac{%
\phi^{\prime}_{n}(\xi)}{\phi_{n} (\xi)},
\end{equation}
in Eq. (70) and suppose that the resulting stationary Schr\"{o}dinger's
equation 
\begin{equation}
\lbrack -\partial _{\xi }^{2}+{\cal V}(\xi )]\phi _{n}(\xi )={\cal E}%
_{n}\phi _{n}(\xi )
\end{equation}
is exactly solvable, where $n$ is a quantum number labelling the eigenvalues
and eigenfunctions. If together with (74) we take 
\begin{equation}
h(\rho )= 2e^{-2c\rho}[{\cal H}(\rho )+{\cal E}_{n}],
\end{equation}
then from Eq. (57) $V_{i}$ are found to be 
\begin{eqnarray}
V_{0} &=& e^{-2c\rho }[{\cal V}(\xi )+{\cal H}(\rho )],  \nonumber \\
V_{1} &=& e^{-2c\rho }[2(\frac{\phi _{n}^{\prime }(\xi )}{\phi _{n}(\xi )}%
)^{2} +2{\cal E}_{n}-{\cal V}(\xi )+{\cal H}(\rho )].  \nonumber
\end{eqnarray}
In that case the separated equations of $H_{0}$ are 
\begin{eqnarray}
\lbrack-\partial _{\xi }^{2}+{\cal V}(\xi )]U_{n}^{0}(\xi )&=&({\cal E}%
_{n}+M)U_{n}^{0}(\xi ), \\
\lbrack-\partial _{\rho }^{2}+{\cal H}(\rho)-e^{2c\rho}E^{0}]R_{n}(\rho )&=&
-({\cal E}_{n}+M)R_{n}(\rho ).
\end{eqnarray}

Let us choose $M$ such that 
\begin{equation}
{\cal E}_{n_{\pm}}= {\cal E}_{n}\pm M
\end{equation}
This amounts to the fact that $\xi$-equation of $H_{0}$ is the same as Eq.
(75). Therefore, $U^{0}_{n}(\xi)=\phi_{n_{+}}(\xi)$ and $E^{0}, R_{n}(\rho)$
must be labelled by $n_{+}$. Accordingly Eq. (78) must be rewritten as 
\begin{equation}
[- \partial _{\rho }^{2}+{\cal H}(\rho)-e^{2c\rho}E^{0}_{n_{+}}]
R_{n_{+}}(\rho ) =-{\cal E}_{n_{+}}R_{n_{+}}(\rho ),
\end{equation}
The eigenvalue equation of $H_{1}$ corresponding to the same $E_{n_{+}}^{0}$
can be separated such that the $\rho $-equation is the same as Eq. (80) and $%
\xi $-equation reads 
\begin{equation}
\lbrack -\partial _{\xi }^{2}+2(\frac{\phi _{n}^{\prime }(\xi )}{\phi
_{n}(\xi )})^{2}-{\cal V}(\xi )]U_{n_{+}}^{1}(\xi )=-{\cal E}%
_{n_{-}}U_{n_{+}}^{1}(\xi ).
\end{equation}
where 
\begin{equation}
U_{n_{+}}^{1}(\xi )={\cal L}U_{n_{+}}^{0}(\xi ) =[-\frac{\phi _{n}^{\prime
}(\xi )}{\phi _{n}(\xi )}+\partial _{\xi }]\phi _{n_{+}}(\xi ).
\end{equation}

The function $\phi_{n}(\xi)$ that generates the transformation is
annihilated by the action of ${\cal L}$, i.e., ${\cal L}\phi_{n}(\xi) =
\{[\phi_{n}^{\prime}(\xi)/\phi_{n}(\xi)] -\partial_{\xi}\}\phi_{n}(\xi)=0$.
Hence, in the case of $M=0$ the function $U_{n}^{1}(\xi)$ corresponding to $%
\phi_{n}(\xi)$ can not be found in this way. But, by referring to a
well-known theorem of the theory of ordinary differential equations $%
U_{n}^{1}$ can be constructed. This theorem says that if $y_{0}(x) $ is a
particular non-trivial solution of the equation $a_{0}(x)y^{\prime
\prime}+a_{1}(x)y^{\prime}+a_{2}(x)y=0$ then the second solution $y_{1}$
linearly independent from $y_{0}$ is given by 
\begin{equation}
y_{1} =y_{0}\int \frac{\exp[-\int\frac{a_{1}(x)}{a_{0}(x)}dx]} {y_{0}^{2}}dx.
\end{equation}
Adopting this theorem to Eq. (75) where $a_{0}=-1$ and $a_{1}=0$ the second
solution linearly independent from $\phi_{n}(\xi)$ is found to be $%
Y(\xi)=\phi_{n}(\xi)\int d\xi/\phi^{2}_{n}(\xi)$. ${\cal L}$ generated by $%
\phi_{n}(\xi)$ applied to $Y(\xi)$ gives $Y_{0}(\xi)={\cal L}%
Y(\xi)=-1/\phi_{n}(\xi)$. Inserting this (as $y_{0}$) into (83) yields the
desired eigenfunction corresponding to $\phi_{n}(\xi)$ 
\begin{eqnarray}
U_{n}^{1}(\xi)=-\frac{1}{\phi_{n}(\xi)}\int \phi_{n}^{2}(\xi)d\xi.  \nonumber
\end{eqnarray}

As a result, changing the eigenfunction of Eq. (75) used to generate the
transformation will lead us to a new eigenvalue problem given by Eq. (81).
In that way a hierarchy of $2D$ isospectral potentials can be constructed.

\section{3D Isospectral Potentials}

In order to find the general solution of Eq. (52) for $n=3$ we firstly
recall the integrability condition (39) and the relation (40). Secondly we
observe that the set $\{ \vec{L}, \vec{c}, \vec{L}{\bf \times} \vec{c} \}$
forms a right-handed (unnormalized) orthogonal moving frame which ``moves"
about fixed direction of $\vec{c}=(c_{1},c_{2},c_{3})=(c_{23},c_{31},c_{12})$%
. By using $x=x_{1}, y=x_{2}, z=x_{3}$ we now introduce the variables 
\begin{eqnarray}
\beta &=& \vec{r}\cdot \vec{c},  \nonumber \\
\gamma &=&\frac{1}{2}\vec{r}\cdot [(\vec{a}+\vec{L}){\bf \times} \vec{c}] = 
\vec{r}\cdot (\vec{a}{\bf \times}\vec{c}) + \frac{1}{2}[(\vec{r}\cdot \vec{c}%
)^{2}-c^{2}r^{2}], \\
\eta &=& \frac{L_{1}}{L_{2}}= \frac{a_{1}+c_{3}y - c_{2}z}{%
a_{2}-c_{3}x+c_{1}z}.  \nonumber
\end{eqnarray}
These obey the following relations 
\begin{eqnarray}
(\vec{L}\cdot \vec{\nabla})\beta &=& \vec{L}\cdot \vec{c}=0, \\
(\vec{L}\cdot \vec{\nabla})\gamma &=& \vec{L}\cdot (\vec{L}{\bf \times} \vec{%
c})=0, \\
(\vec{L}\cdot \vec{\nabla})\eta &=& p(\eta)
\end{eqnarray}
where $p(\eta)$ is a quadratic polynomial in $\eta$: 
\begin{equation}
p(\eta)=c_{3}\frac{L^{2}}{L_{2}^{2}}= \frac{1}{c_{3}}[(c^{2}-c_{2}^{2})%
\eta^{2}+ 2c_{1}c_{2}\eta+(c^{2}-c_{1}^{2})],
\end{equation}
and $c^{2}=c^{2}_{1}+c^{2}_{2}+c^{2}_{3}$. In deriving this we assumed $%
c_{3}\neq 0$ and made use of Eq. (37).

It is now easy to see that, in view of (85) and (86), the general solution
of (52) is $h=h(\beta ,\gamma )$ where $h:R^{2}\rightarrow R$ is an
arbitrary differentiable function. On the other hand, from (53-55) the
general forms of the potentials are 
\begin{equation}
V_{0}=\frac{1}{2}h(\beta ,\gamma )+\frac{V_{-}}{L^{2}},\quad V_{1} = \frac{1%
}{2}h(\beta ,\gamma )+\frac{V_{+}}{L^{2}},
\end{equation}
where 
\begin{eqnarray}
V_{\pm } &=&f^{2}(\eta )\pm p(\eta )f^{\prime }(\eta ), \\
L^{2} &=&a^{2}- 2\gamma .
\end{eqnarray}
Making use of Eqs. (84-87) ${\cal L}$ is found to be 
\begin{eqnarray}
{\cal L}=f(\eta )+(\vec{a}+\vec{r}{\bf \times} \vec{c})\cdot \vec{\nabla}
=f(\eta )+p(\eta )\partial _{\eta }.  \nonumber
\end{eqnarray}

If instead of $\eta =L_{1}/L_{2}$ had we taken one of the variables 
\[
\eta _{2}=\frac{L_{1}}{L_{3}}=\frac{a_{1}+c_{3}y-c_{2}z}{a_{3}+c_{2}x-c_{1}y}%
,\quad \eta _{3}=\frac{L_{2}}{L_{3}}=\frac{a_{2}-c_{3}x+c_{1}z}{%
a_{3}+c_{2}x-c_{1}y}, 
\]
we would have obtained $(\vec{L}\cdot \vec{\nabla})\eta _{j}=p_{j}(\eta
_{j}),j=2,3$ and $V_{\pm }=f_{j}^{2}(\eta _{j})\pm p_{j}(\eta
_{j})f_{j}^{\prime }(\eta _{j})$, where 
\begin{eqnarray}
p_{2}(\eta _{2}) &=&-c_{2}\frac{L^{2}}{L_{3}^{2}}=-\frac{1}{c_{2}}%
[(c^{2}-c_{3}^{2})\eta _{2}^{2}+2c_{1}c_{3}\eta _{2}+(c^{2}-c_{1}^{2})], 
\nonumber \\
p_{3}(\eta _{3}) &=&c_{1}\frac{L^{2}}{L_{3}^{2}}=\frac{1}{c_{1}}%
[(c^{2}-c_{3}^{2})\eta _{3}^{2}+2c_{2}c_{3}\eta _{3}+(c^{2}-c_{2}^{2})]. 
\nonumber
\end{eqnarray}
Without any change in the $\beta ,\gamma $ dependence merely ${\cal L}$
would have been changed as ${\cal L}=f_{j}(\eta _{j})+p_{j}(\eta
_{j})\partial _{\eta _{j}}$.

As an application we again consider the simplest case $V_{0}=0$. Following
an analysis similar to that made in section VI one can easily verify that
the following $3$ different potentials: 
\begin{eqnarray}
V_{1}^{(1)} &=& \frac{2f_{1}^{2}}{L^{2}},  \nonumber \\
V_{1}^{(2)} &=& \frac{2}{L^{2}}[c^{2}+\frac{1}{4}p^{\prime 2}(\eta)], 
\nonumber \\
V_{1}^{(3)} &=& \frac{2}{L^{2}}\{c^{2}+[\frac{1}{2}p^{\prime}(\eta) +\frac{%
p(\eta)}{b_{1}-\eta}]^{2} \},  \nonumber
\end{eqnarray}
are intertwined to $3D$ free motion respectively by 
\begin{eqnarray}
{\cal L}^{(1)}= f_{1}+p(\eta)\partial{\eta}, \quad {\cal L}^{(2)} = \frac{1}{%
2}p^{\prime}(\eta) +p(\eta)\partial{\eta}, \quad {\cal L}^{(3)}= [\frac{1}{2}%
p^{\prime }(\eta) +\frac{p(\eta)}{b_{1}-\eta}] +p(\eta)\partial{\eta}, 
\nonumber
\end{eqnarray}
where $f_{1}= [b_{1}- (cc_{3}^{2})^{-1} \tan^{-1}(p^{\prime
}(\eta)/2c)]^{-1} $ and $b_{1}$ is an integration constant. More generally,
a two parameter family of potentials can be constructed by means of $%
f=(-b)^{1/2}\tan[(-b)^{1/2}(b_{1}+\int d\eta/p(\eta))]$.

We shall now show that in terms of ($\beta ,\gamma ,\eta $) the eigenvalue
equations of $H_{i}$'s accept separation of variables. Starting with 
\begin{eqnarray}
d\beta =\vec{c}\cdot d\vec{r},\quad d\gamma =(\vec{L}{\bf \times} \vec{c}%
)\cdot d\vec{r},\quad d\eta =\frac{c_{3}}{L_{2}^{2}}\vec{L}\cdot d\vec{r},
\end{eqnarray}
one can easily write the differentials $dx,dy,dz$ in terms of $d\beta
,d\gamma ,d\eta $. These are as follow 
\begin{equation}
d\vec{r}=\frac{1}{c^{2}}\vec{c}d\beta +\frac{1}{c^{2}L^{2}}(\vec{L} {\bf %
\times} \vec{c})d\gamma +\frac{L_{2}^{2}}{c_{3}L^{2}}\vec{L}d\eta .
\end{equation}
With the help of these relations the volume form $dV=dx\wedge dy\wedge dz$,
the metric $ds^{2}=d\vec{r}\cdot d\vec{r}$, and $\nabla ^{2}$ are found to
be 
\begin{eqnarray}
dV &=&\frac{1}{c^{2}p(\eta )}d\beta \wedge d\gamma \wedge d\eta , \\
ds^{2} &=&\frac{1}{c^{2}}(d\beta )^{2}+\frac{1}{c^{2}L^{2}}(d\gamma )^{2}+%
\frac{L_{2}^{4}}{c_{3}^{2}L^{2}}(d\eta )^{2}, \\
\nabla ^{2} &=&c^{2}[\partial _{\beta }^{2}+\partial _{\gamma
}(L^{2}\partial _{\gamma })]+\frac{p(\eta )}{L^{2}}\partial _{\eta }[p(\eta
)\partial _{\eta }].
\end{eqnarray}
From Eq. (94) we infer that the Jacobian determinant of the transformation $%
(x,y,z)\rightarrow (\beta ,\gamma ,\eta )$ is $1/c^{2}p(\eta )$. On the
other hand Eq. (95) manifestly shows that the coordinate system $(\beta
,\gamma ,\eta )$ is orthogonal. In virtue of (96) the eigenvalue equation $%
H_{0}\psi ^{0}(\beta ,\gamma ,\eta )=E^{0}\psi ^{0}(\beta ,\gamma ,\eta )$
separates, by taking $\psi ^{0}(\beta ,\gamma ,\eta )=U^{0}(\beta ,\gamma
)R^{0}(\eta )$, as 
\begin{eqnarray}
H_{\beta \gamma }U^{0}(\beta ,\gamma )=MU^{0}(\beta ,\gamma ),\quad H_{\eta
}R^{0}(\eta )=-MR^{0}(\eta ),
\end{eqnarray}
where $M$ is a separation constant and 
\begin{eqnarray}
H_{\beta \gamma } &=&-c^{2}L^{2}[\partial _{\beta }^{2}+\partial _{\gamma
}(L^{2}\partial _{\gamma })]+L^{2}[\frac{1}{2}h(\beta ,\gamma )-E^{0}], \\
H_{\eta } &=&-p(\eta )\partial _{\eta }[p(\eta )\partial _{\eta
}]+f^{2}(\eta )-p(\eta )f^{\prime }(\eta ).
\end{eqnarray}
At this point we will be content with saying that by following the similar
steps as for section VII one can construct hierarchy of $3D$ isospectral
potentials.

Finally we would like to emphasize that the $3D$ potentials we have found
depend on six parameters such that a large number of potentials can be
generated by setting some of them to zero, or, to some particular values.
Possible choices of parameters are represented in the Table I. The
corresponding potentials can be read off from the expressions in the main
text.

\section{CONCLUDING REMARKS}

Main results of this study can be summarized as follows. We have studied a
pair of $nD$ Hamiltonians of potential forms that intertwine by first order
operator ${\cal L}$ and proved that the differential part of ${\cal L}$ is
an element of the Euclidean algebra $e(n)$. These imply that so-intertwined
systems have symmetry operators whose differential parts belong to
enveloping algebra of $e(n)$. The integrability conditions of consistency
equations are dimension dependent and therefore have been considered for
each case separately. The general form of potentials have been specified for 
$n=2,3,4,5 $ where only one linear partial differential equation which
determines the common part of the potential remains unsolved. We have found
the general solution of this equation in cases of $n=2$ and $n=3$.

Three distinctive features of the higher dimensional extension of the
intertwining method are that: (i) The method suggests coordinate systems
which allows us to do the separation of variables and to utilize, in one of
the variable, all the methods of the $1D$ SUSY quantum mechanics. (ii) In
the choice of this variable and ${\cal L}$ itself one has a number of
alternatives increasing with $n$. This fact enlarges the set of available
potentials for each $n\geq 3$. (iii) There exist families of potentials
accepting the same intertwining operator.

$2D$ and $3D$ isospectral potentials we have obtained involve
two arbitrary functions. The former constitute the most
general integrable potentials which admit first order
intertwining. Particular forms of these potentials may be of
special interest for various purposes. Having in mind the projection
techniques which produce exactly solvable lower dimensional problems
from the higher dimensional one-particle problems \cite{Perelomov,Hoppe}
our analysis in section VI-VIII must be continued also for $n=4$
and $n=5$. As is implied by the last example of section VI, it
seems to be possible to investigate connections among the
superintegrable potentials \cite{Evans} as well as to construct
related potentials by repeated Darboux's transformations in
the context of intertwining method. Velocity dependent, stationary
and non-stationary problems \cite{Dorizzi,Sween} can as well
be considered within our approach. Work on $2D$ and $3D$
isospectral potentials which are at the same time superintegrable
is in progress.

\acknowledgements

We thank M. \"{O}nder for a critical reading of this manuscript and for
illuminating discussions. Special thanks are due to A. U. Y\i lmazer and B.
Demircio\u{g}lu for useful conversations. This work was supported in part by
the Scientific and Technical Research Council of Turkey (T\"{U}B\.{I}TAK).

\begin{table}[tbp]
\caption{In 3-dimension the special choices of parameters and corresponding
coordinates. Note that in each case further choices are possible. For
example, in the first three cases $c_{j}$ which does not appear in the first
column can be set to zero and instead of $\protect\eta$, one can also use $%
\protect\eta_{2}$, or, $\protect\eta_{3}$. As a completely different case in
which all $c_{jk}$'s are zero has been presented in section III for any $n>1$%
. }
\label{tab1}
\begin{tabular}{|c|c|c|c|c|}
& $\beta$ & $2\gamma=a^{2}-L^{2}$ & $\eta$ & $p(\eta)$ \\ 
\tableline $a_{1}=0; a_{2}c_{2}+a_{3}c_{3}=0$ & $\vec{r}\cdot \vec{c}$ & $2%
\vec{r}\cdot (\vec{a}{\bf \times} \vec{c})+ (\vec{r}\cdot \vec{c}%
)^{2}-c^{2}r^{2}$ & $\frac{c_{3}y-c_{2}z}{a_{2}-c_{3}x+c_{1}z}$ & $p(\eta)$
\\ 
$a_{2}=0; a_{1}c_{1}+a_{3}c_{3}=0$ & $\vec{r}\cdot \vec{c}$ & $2\vec{r}\cdot
(\vec{a}{\bf \times} \vec{c})+ (\vec{r}\cdot \vec{c})^{2}-c^{2}r^{2}$ & $%
\frac{a_{1}+c_{3}y-c_{2}z}{-c_{3}x+c_{1}z}$ & $p(\eta)$ \\ 
$a_{3}=0; a_{1}c_{1}+a_{2}c_{2}=0$ & $\vec{r}\cdot \vec{c}$ & $2\vec{r}\cdot
(\vec{a}{\bf \times} \vec{c})+ (\vec{r}\cdot \vec{c})^{2}-c^{2}r^{2}$ & $%
\frac{a_{1}+c_{3}y-c_{2}z}{a_{2}-c_{3}x+c_{1}z}$ & $p(\eta)$ \\ 
\tableline $a_{1}=0=a_{2}; c_{3}=0$ & $c_{1}x+c_{2}y$ & $%
2a_{3}(c_{1}y-c_{2}x)+ (c_{1}x+c_{2}y)^{2}-(c_{1}^{2}+c_{2}^{2})r^{2}$ & $%
\frac{-c_{2}z}{a_{3}+c_{2}x-c_{1}y}$ & $p_{2}(\eta_{2})$ \\ 
$a_{1}=0=a_{3}; c_{2}=0$ & $c_{1}x+c_{3}z$ & $2a_{2}(c_{3}x-c_{1}z)+
(c_{1}x+c_{3}z)^{2}-(c_{1}^{2}+c_{3}^{2})r^{2}$ & $\frac{c_{3}y}{%
a_{2}-c_{3}x+c_{1}z}$ & $p(\eta)$ \\ 
$a_{2}=0=a_{3}; c_{1}=0$ & $c_{2}y+c_{3}z$ & $2a_{1}(c_{2}z-c_{3}y)+
(c_{2}y+c_{3}z)^{2}-(c_{2}^{2}+c_{3}^{2})r^{2} $ & $\frac{a_{1}+c_{3}y-c_{2}z%
}{-c_{3}x}$ & $p(\eta)$ \\ 
\tableline $a_{1}=0; c_{2}=0=c_{3}$ & $c_{1}x$ & $2c_{1}(a_{2}y-a_{3}z)
-c_{1}^{2}(y^{2}+z^{2})$ & $\frac{a_{2}+c_{1}z}{a_{3}-c_{1}y}$ & $%
p_{3}(\eta_{3})$ \\ 
$a_{2}=0; c_{1}=0=c_{3}$ & $c_{2}y$ & $%
2c_{2}(-a_{3}x+a_{1}z)-c_{2}^{2}(x^{2}+z^{2})$ & $\frac{a_{1}-c_{2}z}{%
a_{3}+c_{2}x}$ & $p_{2}(\eta_{2})$ \\ 
$a_{3}=0; c_{1}=0=c_{2}$ & $c_{3}z$ & $%
2c_{3}(a_{2}x-a_{1}y)-c_{3}^{2}(x^{2}+y^{2})$ & $\frac{a_{1}+c_{3}y}{%
a_{2}-c_{3}x}$ & $p(\eta)$ \\ 
\tableline $a_{1}=a_{2}=a_{3}=0$ & $\vec{r}\cdot \vec{c}$ & $(\vec{r}\cdot 
\vec{c})^{2}-c^{2}r^{2}$ & $\frac{c_{3}y-c_{2}z}{-c_{3}x+c_{1}z}$ & $p(\eta)$%
\end{tabular}
\end{table}

\end{document}